\documentclass[conference]{IEEEtran}
\IEEEoverridecommandlockouts
\usepackage{cite}
\usepackage{amsmath,amssymb,amsfonts}
\usepackage{algorithmic}
\usepackage{graphicx}
\usepackage{textcomp}
\usepackage{xcolor}
\usepackage{booktabs}
\usepackage{multirow}
\usepackage[T1]{fontenc}
\usepackage{array}
\def\BibTeX{{\rm B\kern-.05em{\sc i\kern-.025em b}\kern-.08em
    T\kern-.1667em\lower.7ex\hbox{E}\kern-.125emX}}
\begin{document}

\title{Code-Level Cost Function Generation for Spatial Image Steganography Using RAG-Enhanced Large Language Models}

\author{
\IEEEauthorblockN{Yige Wang}
\IEEEauthorblockA{\textit{School of Communication \& Information}\\
\textit{Engineering, Shanghai University}\\
Shanghai 200444, China\\
24721152@shu.edu.cn}
\and
\IEEEauthorblockN{Shiqi Yi}
\IEEEauthorblockA{\textit{School of Communication \& Information}\\
\textit{Engineering, Shanghai University}\\
Shanghai 200444, China \\
shiqiyi2025@163.com}
\and
\IEEEauthorblockN{Hanzhou Wu$^*$\thanks{$^*$Author to whom any correspondence should be addressed.}}
\IEEEauthorblockA{\textit{School of Communication \& Information} \\
\textit{Engineering, Shanghai University}\\
Shanghai 200444, China\\
h.wu.phd@ieee.org}}

\maketitle

\begin{abstract}
Designing cost functions of adaptive steganography traditionally requires extensive manual tuning, while deep learning methods lack interpretability. Although large language models (LLMs) offer an automated alternative via evolutionary generation, they often violate domain specific mathematical constraints due to a lack of explicit domain knowledge. To address this problem, we propose a novel evolutionary system focused on exploiting Retrieval-Augmented Generation (RAG) enhanced LLMs for the automatic code-level generation of spatial steganography cost functions. This system incorporates a core Self Evolving RAG (SE-RAG) module, wherein a Code Semantic Signature (CSS) translates procedural code into aligned queries, retrieving explicit guidance from static literature and dynamic experience knowledge bases to steer the LLM generation process. A dedicated feedback mechanism then continuously refines the dynamic knowledge base with successful optimization strategies. Extensive experiments on the BOSSBase and BOWS2 datasets demonstrate that the proposed framework consistently achieves higher steganographic security than existing automatically designed methods, and increases the average code execution rate by 46.3\% while reducing the search cost by 26.1\%, thereby highlighting the effectiveness, efficiency, and potential of combining LLMs with domain-specific knowledge in the field of automatic steganographic algorithm generation.
\end{abstract}

\section{Introduction}
Image steganography conceals secret information within innocuous cover images while minimizing detectable embedding artifacts. Among existing adaptive steganographic approaches, the minimum-distortion framework has achieved remarkable success due to its flexibility and effectiveness, where embedding costs are typically manually assigned according to local image characteristics and optimized using syndrome trellis codes (STCs) \cite{stc}. Although highly effective, designing distortion cost functions still relies heavily on expert knowledge and extensive trial and error \cite{hill,wow,suniward}. Recent deep learning based approaches alleviate manual feature engineering by learning steganographic representations directly from data, at the cost of large training data requirements and limited interpretability. Recently, large language models (LLMs) have begun to be applied to steganographic tasks \cite{prompting}. They have further emerged as a promising alternative for automatic algorithm design through evolutionary code generation \cite{last}. However, without explicit domain knowledge, language models tend to perform unguided evolutionary modifications that may violate mathematical constraints or deviate from effective distortion optimization.

Unlike general code generation, automatic steganographic cost function design is fundamentally a knowledge-intensive task. The objective is not merely to generate executable programs, but to generate mathematically valid distortion functions that satisfy strict steganographic constraints while improving resistance against steganalysis. Such domain-specific knowledge cannot be fully captured by the general-purpose knowledge embedded in LLMs alone. Updating the internal knowledge of LLMs through continual training \cite{ke_continual} or model editing \cite{meng_rome} is computationally expensive and difficult to adapt to the rapidly evolving search process. Retrieval-augmented generation (RAG) \cite{lewis_rag,aigc_survey} has therefore emerged as an effective paradigm for augmenting LLMs with external knowledge without modifying the model parameters. Recent studies have consequently extended RAG through iterative retrieval \cite{iter_retgen}, self-reflective retrieval \cite{self_rag,bi_rag}, and evolutionary retrieval strategies \cite{evor}. Meanwhile, retrieval-enhanced code generation has also demonstrated the benefit of exploiting structured external knowledge for program synthesis \cite{repocoder}. These advances demonstrate that retrieval provides an effective mechanism for incorporating task-specific knowledge into LLM-based generation, making it a promising foundation for automatic steganographic algorithm design.

Motivated by these observations, we propose a knowledge-guided evolutionary framework, which incorporates a self-evolving RAG (SE-RAG) module to provide structured domain guidance throughout the evolutionary process. Specifically, the SE-RAG module jointly exploits a static literature knowledge base and a dynamic experience knowledge base to achieve this goal. We further introduce a code semantic signature (CSS) that bridges procedural implementations and steganographic concepts, enabling semantics-aware retrieval instead of syntax-oriented matching. A feedback-driven knowledge evolution mechanism continuously transforms successful optimization strategies into reusable experience, allowing subsequent generations to evolve under progressively richer domain guidance.

The main contributions are summarized as follows:
\begin{itemize}
    \item We propose a knowledge-driven evolutionary framework for automatic steganographic cost function design, empowered by a novel SE-RAG module that integrates a static literature knowledge base with a dynamic experience knowledge base.
    \item We introduce CSS that bridges the semantic gap between program implementations and steganographic concepts, improving retrieval relevance for cost function generation.
    \item We develop a novel feedback-driven knowledge evolution mechanism that continuously archives successful optimization experiences into the dynamic knowledge base to provide progressively richer domain guidance.
    \item Comprehensive evaluations demonstrate that the algorithms generated by our framework consistently achieve higher steganographic security than existing paradigms. Furthermore, the framework significantly enhances evolutionary efficiency, increasing the average code execution rate by 46.3\% and reducing the search cost by 26.1\%.
\end{itemize}

\begin{figure}[t]
\centerline{\includegraphics[width=\linewidth]{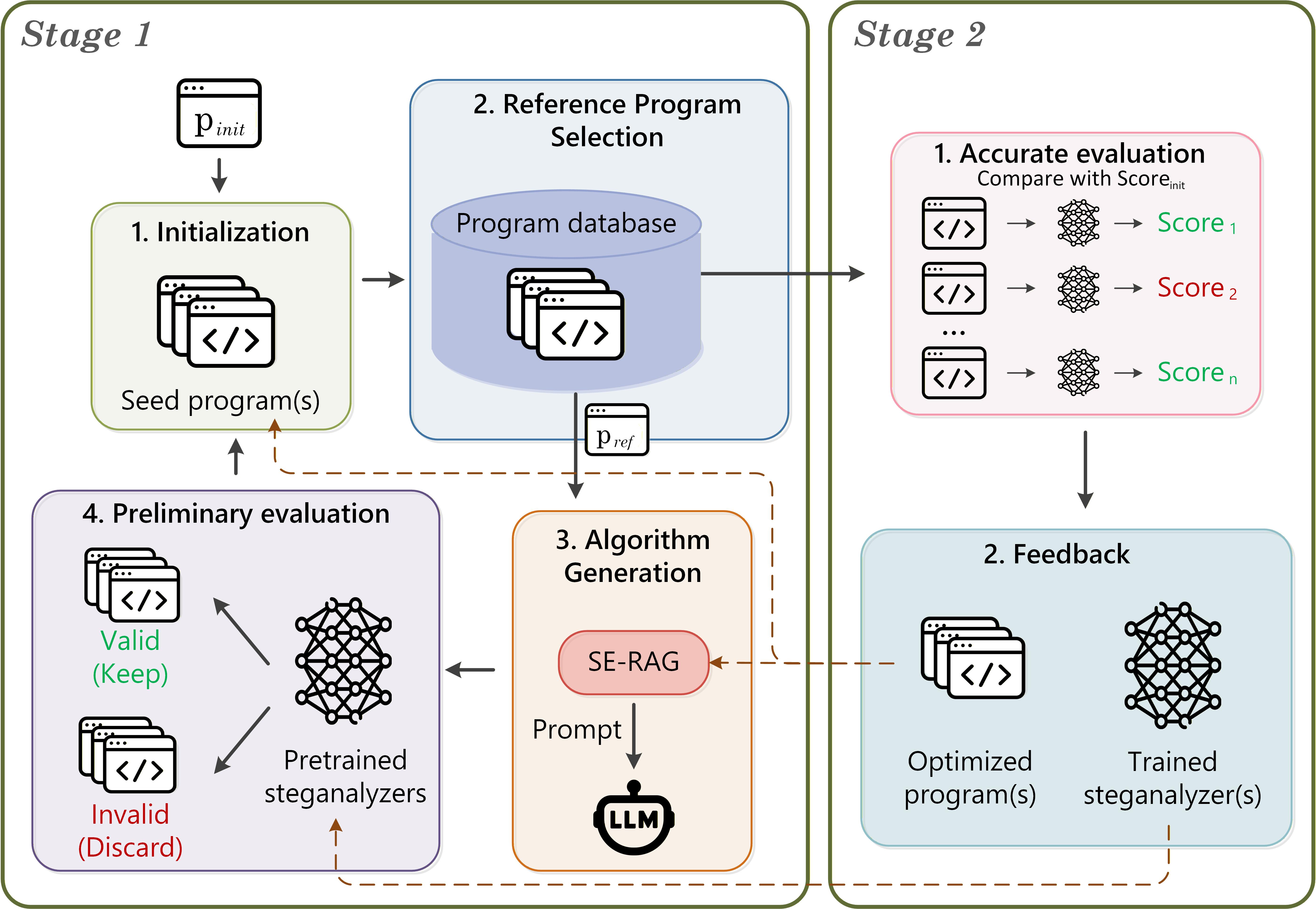}}
\caption{Overview of the proposed framework.}
\label{fig1}
\end{figure}

\section{Proposed Framework}
\subsection{Overall Architecture}
To advance automated cost function design within the minimum-distortion framework, we establish a knowledge-driven evolutionary paradigm. While recent automated frameworks rely on unguided two-stage generation \cite{last}, our architecture fundamentally transforms this process by establishing the SE-RAG module as the core engine for explicit domain guidance. The framework evolves over multiple generations, each comprising cost function generation and evaluation stages. As illustrated in Fig. \ref{fig1}, the process is initialized with a foundational handcrafted steganographic algorithm $p_{\text{init}}$, which acts as the initial program.

Stage 1 focuses on algorithm generation and preliminary evaluation. In each iteration, a reference program $p_{\text{ref}}$ is selected from the program database and fed into the forward process of SE-RAG. Crucially, the CSS of $p_{\text{ref}}$ is extracted to perform separate dual-path retrievals from a static literature knowledge base $\mathcal{K}_{\text{sta}}$ and a dynamic experience knowledge base $\mathcal{K}_{\text{dyn}}^t$, supplying the LLM with classic academic concepts and historically successful optimization paths. Guided by this retrieved context, the LLM optimizes $p_{\text{ref}}$ to generate new candidate algorithms ($p_1, p_2, \dots, p_n$). These variants undergo rapid preliminary evaluation via pretrained steganalyzers to filter out non-executable code. We uniformly utilize the minimum average decision error rate $P_E$ as the security metric:
\begin{equation}
\label{eq:pe}
    P_E = \min_{P_{\text{FA}}} \frac{1}{2}(P_{\text{FA}} + P_{\text{MD}}(P_{\text{FA}})),
\end{equation}
where $P_{\text{FA}}$ and $P_{\text{MD}}$ represent the false alarm and missed detection rates, respectively. Valid programs and their preliminary $P_E$ scores are saved back into the program database.

Stage 2 conducts precise evaluation for candidates passing the preliminary filter to provide feedback. A dedicated steganalyzer is trained for each program to provide a rigorous assessment via the $P_E$ metric defined in Eq.~(\ref{eq:pe}). The results of this accurate evaluation update the pretrained steganalyzer pool in Stage 1, and the optimized programs are added to the seed program pool for subsequent generations. Simultaneously, the evaluation results leverage the backward process of SE-RAG to update the dynamic experience knowledge base $\mathcal{K}_{\text{dyn}}^t$ based on the evaluation results, accumulating effective empirical paths for continuous improvement.

\begin{figure}[t]
\centerline{\includegraphics[width=\linewidth]{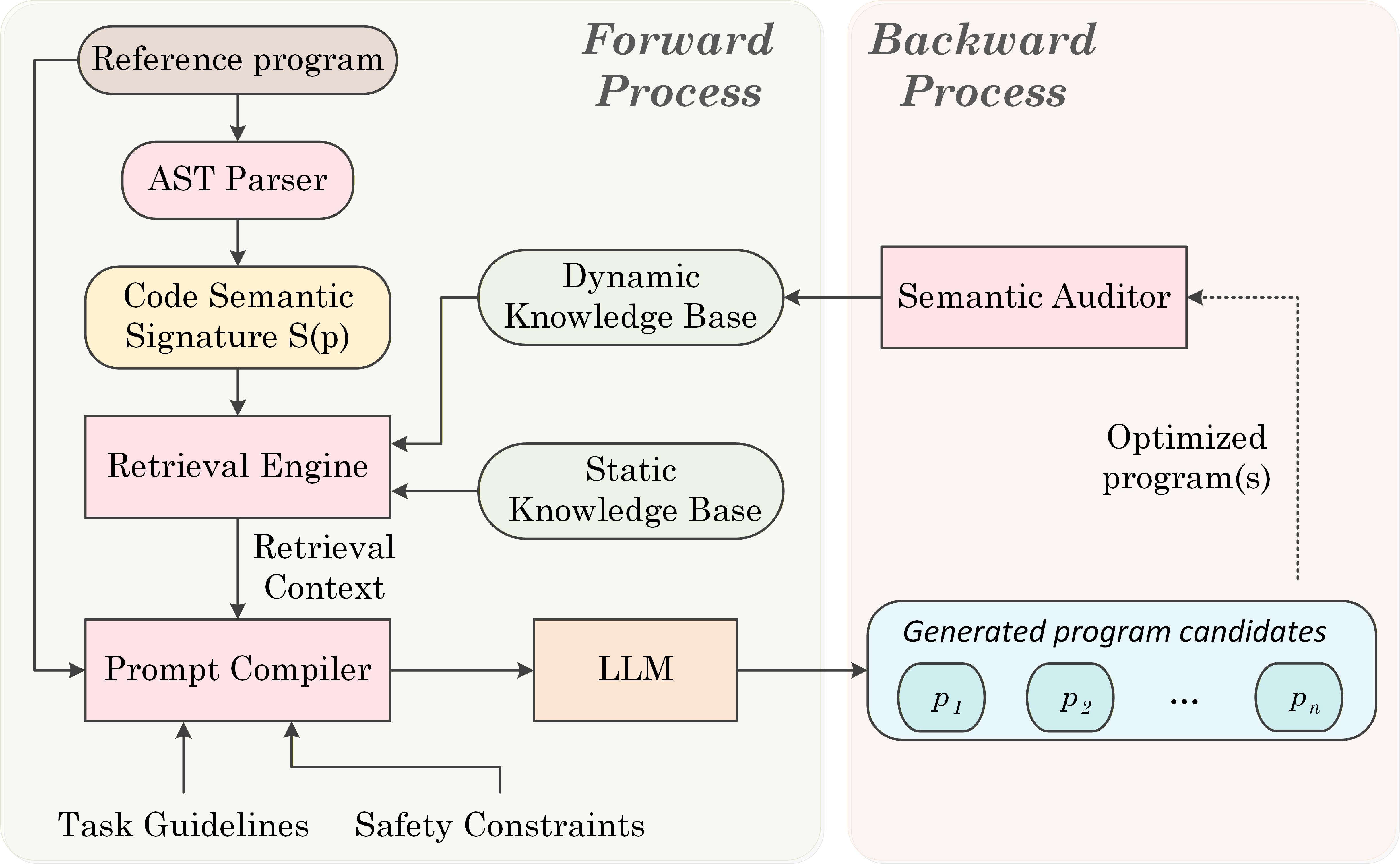}}
\caption{The internal pipeline of the SE-RAG module.}
\label{fig2}
\end{figure}

The overall workflow is illustrated in Fig. \ref{fig1}, and the internal pipeline of the SE-RAG module is detailed in Fig. \ref{fig2}. Section \ref{subsec: Forward Feedback Process} describes the forward generation process, while Section \ref{subsec: Backward Feedback Process} outlines the backward feedback loop.

\subsection{SE-RAG: Forward Generation Process}
\label{subsec: Forward Feedback Process}

The forward generation process transitions the LLM from a blind code generator into a knowledge-informed algorithm optimizer via domain knowledge injection. To achieve effective retrieval, relying directly on the raw reference program $p_{\text{ref}}$ as a query is suboptimal due to the conceptual gap between procedural implementations and steganographic semantics. To bridge this gap, as illustrated in Fig. \ref{fig2}, we introduce the CSS mechanism, where $p_{\text{ref}}$ is parsed by an AST parser. This parser maps specific Python operations to high-level steganographic semantic concepts based on the mapping rules standardized in Table \ref{tab:semantic_mapping}, sequentially connecting them into an ordered semantic sequence $S(p)$.

\begin{table}[t]
\centering
\caption{Mapping from Python Code Patterns to Semantic Concepts}
\label{tab:semantic_mapping}
\begin{tabular}{@{}ll@{}}
\toprule
Raw Python Operation & Semantic Concept \\ \midrule
\texttt{convolve} + \texttt{reciprocal} & Texture Detection \\
\texttt{convolve after reciprocal} & Spatial Cost Smoothing \\
\texttt{roll} & Edge Offset \\
\texttt{slice} & Spatial Cropping \\
\texttt{>= 255} / \texttt{<= 0} & Boundary Value Constraints \\
\texttt{1 / x} & Cost Initialization Inverse \\
\bottomrule
\end{tabular}
\end{table}
Prior to retrieval, literature documents and historically successful rules are partitioned into semantic text chunks $k$. These chunks populate two repositories, namely, the Static Knowledge Base $\mathcal{K}_{\text{sta}}$ indexing fundamental reference papers, and the Dynamic Knowledge Base $\mathcal{K}_{\text{dyn}}^t$, which continuously expands at each evolutionary iteration $t$ to store concise optimization summaries. Formally, each repository $\mathcal{K}_i \in \{\mathcal{K}_{\text{sta}}, \mathcal{K}_{\text{dyn}}^t\}$ is defined as a set of constituent chunks: $\mathcal{K}_i = \{k_1, k_2, \dots, k_{|\mathcal{K}_i|}\}$. All chunks are processed by a Bi-Encoder embedding model $M_{\text{BE}}$ and stored alongside their continuous vector representations.

During the retrieval phase, the obtained semantic signature formulates the query $q = S(p_{\text{ref}})$. A parallel two-stage retrieve-and-rerank pipeline is independently applied to each repository. First, $M_{\text{BE}}$ projects $q$ into the latent space for a rapid vector screening, where $q$ is compared against every chunk $k \in \mathcal{K}_i$ to calculate the preliminary similarity score $s_1(q, k)$ via cosine similarity:
\begin{equation}
    s_1(q, k) = \frac{M_{\text{BE}}(q) \cdot M_{\text{BE}}(k)}{\|M_{\text{BE}}(q)\| \|M_{\text{BE}}(k)\|}.
\end{equation}
This screens the search space to retain a highly relevant matching pool $\mathcal{M}_{i,q}$. Second, the matching pools are forwarded to a Cross-Encoder model $M_{\text{CE}}$ for precise reranking to capture fine-grained cross-attention. The final ranking score is computed as:
\begin{equation}
    s_2(q, k) = M_{\text{CE}}(q, k), \quad \forall k \in \mathcal{M}_{i,q}.
\end{equation}
Finally, a prompt compiler systematically aggregates the acquired information into a highly structured prompt to govern the LLM code synthesis. Instead of a simple concatenation, the compilation process utilizes a sequential integration function $\Phi$ to combine four essential components: First, the expert task guidelines $\mathcal{T}$, which assign a domain expert persona to the language model and explicitly mandate structural algorithmic innovations, such as novel residual sources or nonadditive fusion rules, while strictly prohibiting superficial parameter tuning; Second, the retrieved context $\mathcal{C}_q$, which supplies explicit domain guidance and historical optimization strategies; Third, the reference program $p_{\text{ref}}$ undergoing evolution; and Fourth, the rigorous safety constraints $\mathcal{R}$, which enforce deterministic execution, restrict external dependencies to standard numerical packages, and define rigorous mathematical boundaries to satisfy necessary steganographic security constraints. This compilation operation is formally defined as:
\begin{equation}
    \mathcal{P} = \Phi(\mathcal{T}, \mathcal{C}_q, p_{\text{ref}}, \mathcal{R}).
\end{equation}

The resulting knowledge-injected prompt is then delivered to the LLM to generate the new batch of program candidates.

\subsection{SE-RAG: Backward Feedback Loop}
\label{subsec: Backward Feedback Process}

To prevent the evolutionary process from stalling or repeating past optimization errors, the proposed framework incorporates a backward semantic knowledge extraction mechanism driven by a semantic auditor. The core purpose of this component is to transform fortuitous performance improvements into interpretable steganographic rules, which are subsequently leveraged to guide future mutation directions.

As illustrated in the backward process of Fig. \ref{fig2}, during the accurate evaluation phase, a dedicated steganalysis tool evaluates the optimized program $p_c$ to obtain its security score $P_E(p_c)$, as defined in Eq.~(\ref{eq:pe}). If this score exceeds that of the initial global baseline, i.e., $P_E(p_c) > P_E(p_{\text{init}})$, the knowledge extraction mechanism inside the Semantic Auditor is triggered. The framework employs an independent LLM summarizer within the auditor to perform a detailed comparative analysis between $p_c$ and $p_{\text{init}}$. By examining the algorithmic structures and semantic variations, the summarizer extracts and summarizes the specific methodological modifications that lead to the enhanced security. The empirical insights generated during the current iteration $t$ form a set:
\begin{equation}
    E_{\text{new}}^t = \{e^{(1)}, e^{(2)}, \dots\}.
\end{equation}

Consistent with the knowledge representation established in Section \ref{subsec: Forward Feedback Process}, each empirical experience $e \in E_{\text{new}}^t$ is partitioned into semantic text chunks, embedded into vector representations, and structurally archived. Let $\mathcal{K}_{\text{new}}^t = \{k_1, k_2, \dots\}$ denote the set of all semantic chunks derived from $E_{\text{new}}^t$. The dynamic knowledge base is updated via a direct set union operation:
\begin{equation}
    \mathcal{K}_{\text{dyn}}^t = \mathcal{K}_{\text{dyn}}^{t-1} \cup\mathcal{K}_{\text{new}}^t,
\end{equation}
where $\mathcal{K}_{\text{dyn}}^0 = \emptyset$. 

Integrating this structured knowledge back into the loop establishes a sustainable self-evolving mechanism. As evolution progresses, the dynamic knowledge base $\mathcal{K}_{\text{dyn}}^t$ continuously accumulates historically successful design rules. Consequently, when the next iteration performs forward retrieval, the retrieval engine draws directional guidance from both the foundational static knowledge base $\mathcal{K}_{\text{sta}}$ and the updated dynamic knowledge base $\mathcal{K}_{\text{dyn}}^{t}$, drastically accelerating convergence toward highly secure steganographic cost functions.

\section{Experimental Setup}
Evolutionary generation and evaluation are conducted on BOSSBase v1.01 \cite{BOSSBASE}, with optimal algorithms further tested on BOWS2 \cite{BOWS2} for generalization. Both datasets contain 10,000 $256 \times 256$ grayscale images, strictly partitioned at a 5:1:4 ratio for paired cover-stego training, validation, and testing. All algorithms are implemented in Python via an embedding simulator.

OpenAI GPT-4o \cite{GPT} serves as the core LLM for both code generation and backward experience summarization. Knowledge bases are managed via Chroma DB using chunks of 1,000 characters with a stride of 800. We employ the Bi-Encoder \texttt{all-MiniLM-L6-v2} \cite{MiniLM} for dense retrieval and the Cross-Encoder \texttt{ms-marco-MiniLM-L-6-v2} for reranking. The static repository indexes 228 classical steganography papers. To balance noise and length disparities, the retrieve-and-rerank pipeline filters the top 50 static and 20 dynamic chunks down to $N_{\text{sta}} = 3$ and $N_{\text{dyn}} = 5$, respectively. The evolution is initialized with WOW \cite{wow}, HILL \cite{hill}, and S-UNIWARD \cite{suniward}. Adopting hyperparameters from \cite{last}, each prompt directs the LLM to synthesize four candidate variants per iteration.

For search efficiency, evolution executes exclusively at the highest payload of 0.4 bpp using a fully trained SRNet \cite{srnet}. The discovered algorithms are then evaluated across lower payloads from 0.3 to 0.1 bpp. The SRNet training parameters and progressive fine-tuning strategy strictly follow the configurations in \cite{last} and the original implementation \cite{srnet}.

\section{Results and Analysis}
\subsection{Steganographic Security}

\begin{table}[t]
\caption{Detection error of original algorithms and the best evolved algorithms discovered by different automatic design frameworks under different embedding payloads on BOSSBase (higher values indicate stronger security).}
\label{tab:bossbase_overall_performance}
\centering
\begin{tabular}{>{\centering\arraybackslash}p{1.3cm} >{\centering\arraybackslash}p{1.3cm} c c c c}
\hline
    
\multirow{2}{*}{Algorithm}
& \multirow{2}{*}{Framework}
& \multicolumn{4}{c}{Embedding Payload (bpp)} \\
\cline{3-6}
& & 0.4 & 0.3 & 0.2 & 0.1 \\
\hline

\multirow{3}{*}{HILL}
& Original & 0.1574 & 0.2063 & 0.2596 & 0.3449 \\
& Ref.~\cite{last} & 0.1654 & 0.2241 & 0.2700 & 0.3559 \\
& Ours & \textbf{0.2036} & \textbf{0.2433} & \textbf{0.3064} & \textbf{0.3878} \\
\hline

\multirow{3}{*}{WOW}
& Original & 0.1053 & 0.1361 & 0.1800 & 0.2640 \\
& Ref.~\cite{last} & 0.1385 & 0.1695 & 0.2180 & 0.2911 \\
& Ours & \textbf{0.1419} & \textbf{0.1711} & \textbf{0.2249} & \textbf{0.3186} \\
\hline

\multirow{3}{*}{S-UNIWARD}
& Original & 0.1139 & 0.1573 & 0.2169 & 0.3245 \\
& Ref.~\cite{last} & 0.1245 & 0.1709 & 0.2309 & 0.3376 \\
& Ours & \textbf{0.1588} & \textbf{0.2033} & \textbf{0.2588} & \textbf{0.3608} \\
\hline
\end{tabular}
\end{table}

\begin{table}[t]
\caption{Detection error of original algorithms and the best evolved algorithms discovered by different automatic design frameworks under different embedding payloads on BOWS2 (higher values indicate stronger security).}
\label{tab:bows2_overall_performance}
\centering
\begin{tabular}{>{\centering\arraybackslash}p{1.3cm} >{\centering\arraybackslash}p{1.3cm} c c c c}
\hline

\multirow{2}{*}{Algorithm}
& \multirow{2}{*}{Framework}
& \multicolumn{4}{c}{Embedding Payload (bpp)} \\
\cline{3-6}
& & 0.4 & 0.3 & 0.2 & 0.1 \\
\hline

\multirow{3}{*}{HILL}
& Original & 0.2515 & 0.2978 & 0.3601 & 0.4279 \\
& Ref.~\cite{last} & 0.2738 & 0.3198 & 0.3739 & 0.4450 \\
& Ours & \textbf{0.3476} & \textbf{0.4101} & \textbf{0.4584} & \textbf{0.4848} \\
\hline

\multirow{3}{*}{WOW}
& Original & 0.1605 & 0.1974 & 0.2680 & 0.3644 \\
& Ref.~\cite{last} & 0.2189 & 0.2604 & 0.3284 & 0.4093 \\
& Ours & \textbf{0.2344} & \textbf{0.2811} & \textbf{0.3515} & \textbf{0.4451} \\
\hline

\multirow{3}{*}{S-UNIWARD}
& Original & 0.1931 & 0.2536 & 0.3174 & 0.4229 \\
& Ref.~\cite{last} & 0.2194 & 0.2679 & 0.3499 & 0.4378 \\
& Ours & \textbf{0.2541} & \textbf{0.2999} & \textbf{0.3620} & \textbf{0.4510} \\
\hline
\end{tabular}
\end{table}

To evaluate the effectiveness of the proposed framework, we compare the steganographic security of the original handcrafted algorithms and the best algorithms generated by our framework with those produced by the existing LLM-based approach \cite{last}. HILL \cite{hill}, WOW \cite{wow}, and S-UNIWARD \cite{suniward} are adopted as the starting algorithms for the evolutionary search process in both frameworks. Security is quantified by the detection error rate $P_E$ defined in Eq.~(\ref{eq:pe}), which is evaluated using dedicated SRNet \cite{srnet} steganalyzers trained specifically for each evaluated algorithm at each embedding payload.

As reported in Table~\ref{tab:bossbase_overall_performance} and Table~\ref{tab:bows2_overall_performance}, while the algorithms produced by both automated paradigms successfully enhance the security of their handcrafted counterparts, those generated by our framework consistently achieve the highest $P_E$ across varying payloads on both BOSSBase and BOWS2. By continuously outperforming the baseline across divergent steganographic paradigms, this framework-level comparison systematically demonstrates the superior optimization capacity of our evolutionary framework incorporating the SE-RAG engine, proving that the security gains stem from methodological advancements rather than stochastic fluctuations during the evolutionary search. Consequently, rather than relying solely on the implicit knowledge of the language model, our framework explicitly incorporates domain-specific knowledge into the evolutionary process via the SE-RAG module, enabling a more meaningful exploration of the program space and consistently producing higher-quality steganographic algorithms.

\subsection{Evolution Efficiency Analysis}

\begin{table}[t]
\caption{Evolution efficiency comparison between different automatic design frameworks.}
\label{tab:efficiency}
\centering
\begin{tabular}{llccc}
\toprule
\multirow{2}{*}{Metric} & \multirow{2}{*}{Framework} & \multicolumn{3}{c}{Initial Algorithm} \\ \cmidrule(lr){3-5}
& & HILL & WOW & S-UNIWARD \\ \midrule

\multirow{2}{*}{Execution Rate ($\uparrow$)} 
& Ref.~\cite{last} & 35.0\% & 36.3\% & 20.0\% \\
& Ours        & \textbf{57.5\%} & \textbf{54.2\%} & \textbf{21.9\%} \\ \midrule

\multirow{2}{*}{Target Iterations ($\downarrow$)} 
& Ref.~\cite{last} & 100 & 160 & 240 \\
& Ours        & \textbf{80} & \textbf{120} & \textbf{160} \\ 
\bottomrule
\end{tabular}
\end{table}

To further evaluate the proposed framework, we analyze its evolutionary efficiency in terms of code executability, defined as the percentage of successfully executed candidate algorithms, and search efficiency, representing the number of generated algorithms required to discover the best-performing solution.

As shown in Table~\ref{tab:efficiency}, the proposed framework improves the overall code execution rate across all three initial algorithms, increasing the average execution rate from 30.4\% to 44.5\%, which corresponds to a relative improvement of 46.3\%. Although the execution rate enhancement for the highly constrained S-UNIWARD is marginal, the overall search efficiency is consistently optimized. Specifically, our framework requires substantially fewer generated algorithms to discover the best-performing solution across all baselines, reducing the overall search cost by an average of 26.1\%. Collectively, despite the inherent challenges in optimizing complex spatial steganographic models like S-UNIWARD, these results demonstrate that our framework enables more reliable candidate generation and highly efficient optimization. This substantially reduces the computational cost of automatic steganographic algorithm design, highlighting its practical applicability.

\subsection{Analysis of Evolved Cost Functions}
To explore the effectiveness of our evolutionary system, we present the evolved cost functions derived from HILL, WOW, and S-UNIWARD, detailing the specific modifications made to each algorithm.

\subsubsection{Evolved HILL}
The original HILL algorithm estimates distortion via a high-pass residual $R_H = H_1 \otimes C$, where $C \in \mathbb{R}^{M \times N}$ is the cover image, $\otimes$ denotes convolution, and $H_1$ is the $3 \times 3$ Ker-Bohme (KB) filter. It formulates the suitability map and cost as:
\begin{equation}
\xi_{\text{orig}} = |R_H| \otimes L_1,~\rho_{\text{orig}} = \frac{1}{\xi_{\text{orig}} + \epsilon} \otimes L_2,
\end{equation}
where $L_1$ and $L_2$ are $3 \times 3$ and $15 \times 15$ mean filters, respectively, and $\epsilon = 10^{-8}$ ensures numerical stability.

The evolved algorithm refines this by introducing a median residual $R_M = C - \text{Med}_{3 \times 3}(C)$, where $\text{Med}_{3 \times 3}(\cdot)$ denotes a $3 \times 3$ local median filter. These residuals are fused into a joint response:
\begin{equation}
    R_F = \sqrt{|R_H \cdot R_M|},
\end{equation}
where $\cdot$, $|\cdot|$, and $\sqrt{\cdot}$ represent element-wise multiplication, absolute value, and square root. To measure texture complexity, the local variance $V(C)_{i,j} = \text{Var}\left( \Omega_{3 \times 3}(C_{i,j}) \right)$ within a $3 \times 3$ neighborhood $\Omega_{3 \times 3}$ at each coordinate $(i,j)$ is computed, and immediately truncated into a variance gating factor $\tilde{V}(C)_{i,j} = \text{clip}(V(C)_{i,j}, \epsilon, 5)$, where $\text{clip}(x, a, b)$ restricts $x$ within $[a, b]$. Finally, the evolved algorithm constructs a refined suitability map $\xi$ and maps it directly to the local embedding cost $\rho_{i,j}$:
\begin{equation}
\xi = R_F \otimes L_1,~\rho_{i,j} = \frac{1}{(\xi_{i,j} + \epsilon) \cdot \tilde{V}(C)_{i,j}}.
\end{equation}

To enforce physical boundaries and numerical stability, valid costs are lower-bounded by $\epsilon$. Extreme numerical values and invalid boundary modifications are strictly penalized with an infinite wet cost.

\subsubsection{Evolved WOW}
The original WOW employs Daubechies 8-tap (DB-8) low-pass ($\mathbf{h}$) and high-pass ($\mathbf{g}$) filters to construct three directional filters via outer products to represent the LH, HL, and HH subbands: $K^{(0)} = \mathbf{h} \mathbf{g}^\top$, $K^{(1)} = \mathbf{g} \mathbf{h}^\top$, and $K^{(2)} = \mathbf{g} \mathbf{g}^\top$. The embedding suitability maps quantify local texture complexity by measuring weighted absolute residual differences:
\begin{equation}
    \xi_{\text{orig}}^{(k)} = |C \otimes K^{(k)}| \odot |K^{(k)}|, \quad k \in \{0, 1, 2\},
\end{equation}
where $\odot$ represents mirror-padded correlation. Pixels with small directional residuals are highly predictable, thus receiving higher embedding costs to minimize detectability. The original cost aggregates the reciprocal of these maps:
\begin{equation}
    \rho_{\text{orig}} = \frac{1}{\xi_{\text{orig}}^{(0)}} + \frac{1}{\xi_{\text{orig}}^{(1)}} + \frac{1}{\xi_{\text{orig}}^{(2)}}.
\end{equation}

Replacing conventional hand-crafted wavelets, our evolutionary system autonomously discovers an asymmetric 9-tap filter pair, denoted as $\mathbf{h}_e$ and $\mathbf{g}_e$. Substituting the DB-8 vectors, it reconstructs the three standard subbands and introduces a fourth low-frequency structural subband $K^{(3)} = \mathbf{h}_e \mathbf{h}_e^\top$ to capture local structural consistency. The expanded suitability maps use the identical formulation:
\begin{equation}
    \xi^{(m)} = |C \otimes K^{(m)}| \odot |K^{(m)}|, \quad m \in \{0, 1, 2, 3\}.
\end{equation}

The evolved algorithm replaces static reciprocal aggregation with a pixel-wise dynamic fusion mechanism. Specifically, the directional discrepancy is first evaluated using the channel-wise variance $V(\xi)_{i,j} = \text{Var}\left( \xi^{(0)}_{i,j}, \xi^{(1)}_{i,j}, \xi^{(2)}_{i,j}, \xi^{(3)}_{i,j} \right)$, which is scaled via global normalization to yield $\tilde{V}_{\xi}(i,j) = V(\xi)_{i,j} / (\max(V(\xi)) + \epsilon)$. A dynamic pooling parameter $P_{i,j}$ is then generated to locally adjust the fusion intensity:
\begin{equation}
    P_{i,j} = -0.7 + 0.3 \tilde{V}_{\xi}(i,j).
\end{equation}

To strictly ensure mathematical validity across spatially varying parameters, the aggregation is performed at each pixel by defining a joint suitability pool $\Psi_{i,j}$:
\begin{equation}
\Psi_{i,j} = \sum_{m=0}^{3} \left(\xi^{(m)}_{i,j}\right)^{P_{i,j}},~\rho_{i,j} = \Psi_{i,j}^{-1/P_{i,j}}.
\end{equation}

Cost bounding and numerical stabilization follow the identical protocols established in the evolved HILL algorithm.

\subsubsection{Evolved S-UNIWARD}
The original S-UNIWARD evaluates distortion using directional relative change profiles. Following WOW, three DB-8 directional filters, denoted by $K^{(k)}$ for $k \in \{0,1,2\}$, are employed to compute the directional suitability maps:
\begin{equation}
    \xi_{\text{orig}}^{(k)} = \frac{1}{|C \otimes K^{(k)}| + \sigma} \odot |K^{(k)}|, \quad k \in \{0, 1, 2\},
\end{equation}
where $\sigma = 1$ is a stabilization constant. The overall embedding cost is then obtained by aggregating the three directional suitability maps:
\begin{equation}
    \rho_{\text{orig}} = \xi_{\text{orig}}^{(0)} + \xi_{\text{orig}}^{(1)} + \xi_{\text{orig}}^{(2)}.
\end{equation}

The evolved algorithm expands this design by augmenting the filter bank. Retaining the classic DB-8 filters, it introduces an approximate horizontal gradient filter $K^{(3)} = [-0.5, 0, 0.5]$ and its vertical counterpart $K^{(4)} = (K^{(3)})^\top$ to evaluate local boundary discontinuities. To ensure optimal compliance with the physical properties of different residuals, the expanded suitability maps are extracted through divergent smoothing strategies. The gradient subbands employ a $5 \times 5$ local median filter, while the first three subbands retain their original strategies:
\begin{equation}
    \xi^{(m)} = 
    \begin{cases} 
    \frac{1}{|C \otimes K^{(m)}| + \sigma} \odot |K^{(m)}|, & m \in \{0, 1, 2\},\\ 
    \text{Med}_{5 \times 5}\left( \frac{1}{|C \otimes K^{(m)}| + \sigma} \right), & m \in \{3, 4\}.
    \end{cases}
\end{equation}

A geometric mean aggregation supersedes the original linear additive summation in this expanded design. The final embedding cost is therefore formulated by explicitly evaluating the joint product of all five suitability tracks at each pixel:
\begin{equation}
    \rho_{i,j} = \left( \xi^{(0)}_{i,j} \cdot \xi^{(1)}_{i,j} \cdot \xi^{(2)}_{i,j} \cdot \xi^{(3)}_{i,j} \cdot \xi^{(4)}_{i,j} \right)^{1/5}.
\end{equation}
As with the preceding algorithms, physical boundaries and extreme numerical values are regulated through identical constraint mechanisms.

\section{CONCLUSION}
In this paper, we propose a knowledge-based evolutionary framework for the generation of code-level cost functions in spatial image steganography. To overcome the lack of domain-specific guidance in existing LLM-based automated design, our framework is driven by a central SE-RAG module. Within this engine, CSS is utilized to bridge the semantic gap between procedural implementations and steganographic principles, enabling the continuous and dynamic integration of explicit guidance from both static literature and historically successful optimization experiences.

Experimental results on BOSSBase and BOWS2 demonstrate that our framework consistently achieves higher steganographic security while improving code executability and evolutionary efficiency over existing methods. Crucially, by explicitly injecting structured domain knowledge into the generation loop, this architecture successfully transforms the LLM from a blind code synthesizer into a directed algorithm optimizer. This knowledge-driven guidance effectively prevents mathematically invalid mutations and accelerates convergence toward highly secure cost functions. Future research will focus on expanding the analytical capabilities of the semantic auditor, broadening the evaluation framework across a wider spectrum of steganalyzers to validate universal algorithmic resistance, and adapting this offline knowledge-driven architecture to broader media forensics applications.

\section*{Acknowledgments}
This research was supported by the Science and Technology Commission of Shanghai Municipality (STCSM) under grant 24ZR1424000, National Natural Science Foundation of China (NSFC) under grant U23B2023, and the Nanning ``Yong Jiang'' Program under grant RC20250102.

\vspace{12pt}

\end{document}